\documentclass[twocolumn,showpacs,preprintnumbers,amsmath,amssymb,showkeys,aps,prb,superscriptaddress,10pt]{revtex4-1} 
\usepackage{lipsum}
\usepackage{printlen}
\usepackage{layouts}
\usepackage[scaled]{helvet}
\usepackage[T1]{fontenc}

\usepackage[english]{babel}
\usepackage{graphicx}
\usepackage{mathrsfs}
\usepackage[dvipsnames]{xcolor}
\usepackage[caption=false]{subfig}
\usepackage{float}
\usepackage[section]{placeins}
\usepackage{numprint}
\usepackage{todonotes}
\usetikzlibrary{matrix}
\usepackage{hyperref}
\usepackage[margin=1in]{geometry} 

\usepackage{soul}
\usepackage{array}
\usepackage{booktabs}
\usepackage[referable]{threeparttablex}
\usepackage{adjustbox}
\usepackage{tabularx}
\usepackage{multirow}
\usepackage{makecell}

\usepackage{tikz-3dplot}
\usetikzlibrary{decorations.markings,arrows,arrows.meta, decorations.pathmorphing,backgrounds, positioning, fit, shapes.geometric,shapes,bending,shadows,shadings,shapes.misc,spy}
\usetikzlibrary{calc}
\usetikzlibrary{math}
\usepackage[version=4]{mhchem}

\begin{document}
\widetext
\title[Article Title]{Dynamical manipulation of polar topologies from acoustic phonon excitations}
\author{Louis Bastogne}
\thanks{These authors contributed equally}
\affiliation{Theoretical Materials Physics, Q-MAT, Universit\'e de Li\`ege, Belgium}%
\author{Fernando G\'{o}mez-Ortiz}
\thanks{These authors contributed equally}
\affiliation{Theoretical Materials Physics, Q-MAT, Universit\'e de Li\`ege, Belgium}%
\author{Sriram Anand}
\affiliation{Theoretical Materials Physics, Q-MAT, Universit\'e de Li\`ege, Belgium}%
\author{Philippe Ghosez}
\email{Philippe.Ghosez@uliege.be} %
\affiliation{Theoretical Materials Physics, Q-MAT, Universit\'e de Li\`ege, Belgium}%
\date{\today}

\begin{abstract}
\noindent
\hspace{-0.85in}
\begin{minipage}[t]{0.63\textwidth}
\vspace{0.05in}
    \noindent\textbf{Abstract:}
Since the recent discovery of polar topologies, a recurrent question has been in the way to remotely tune them. Many efforts have focused on the pumping of polar optical phonons from optical methods but with limited success, as only switching between specific phases has been achieved so far. Additionally, the correlation between optical pulse characteristics and the resulting phase remains poorly understood. Here, we propose an alternative approach and demonstrate the deterministic and dynamical tailoring of polar topologies using instead acoustic phonon excitations. Our second-principles simulations reveal that by pumping  specific longitudinal and transverse acoustic phonons, various topological textures can be induced in materials like BaTiO$_\mathrm{3}$ or PbTiO$_\mathrm{3}$. This method leverages the strong coupling between polarization and strain in these materials, enabling predictable and dynamical control of polar patterns. Our findings open up an alternative possibility for the manipulation of polar textures, inaugurating a promising research direction.
\end{minipage}
\hfill
\begin{minipage}[t]{0.34\textwidth}
\vspace{0.3in}
    \hspace{0.3cm}
    \includegraphics[width=7cm, height=3.78cm]{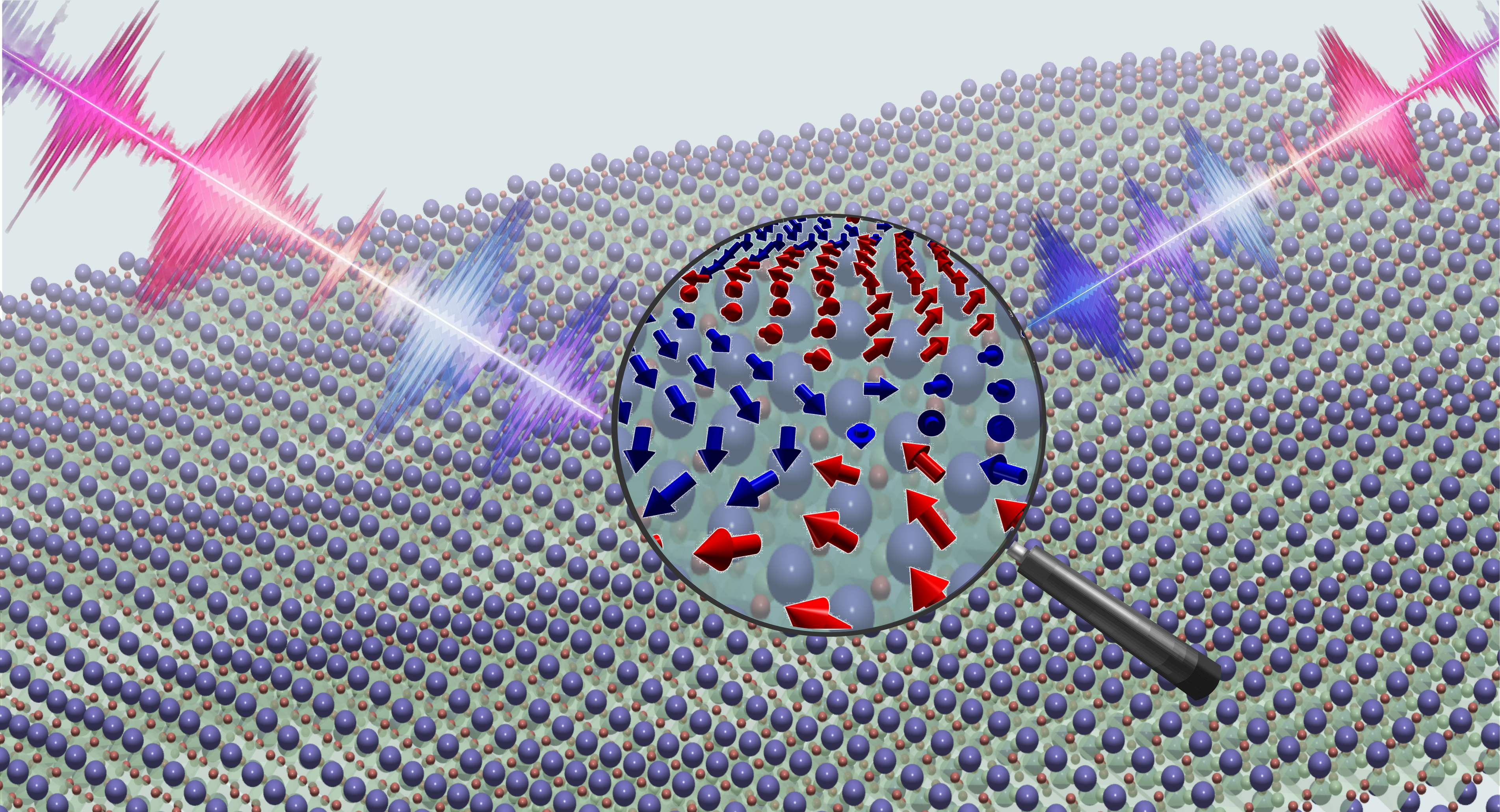}
\end{minipage}
\end{abstract}
\maketitle{}

Ferroelectric materials tend to form uniform domains, which are regions of space characterized by a homogeneous polarization. 
However, more complex configurations have been discovered in recent years, especially in ferroelectric thin films and ferroelectric/dielectric junctions~\cite{Fong-04,Junquera-23}. 
In these cases, the development of a homogeneous polarization state is precluded by the electrostatic penalty associated to the bound charges at the interfaces~\cite{Stepanovich-05} and a compromise between electrostatic, elastic and gradient energies must be achieved. 
As a result, a plethora of different phases have been observed and predicted such as flux-closure domains~\cite{Streiffer-02,Zubko-10,Tang-15}, polar vortices~\cite{Yadav-16}, skyrmions~\cite{Das-19}, merons~\cite{Wang-20}, hopfions~\cite{Lukyanchuk-20} or the so-called supercrystals~\cite{Stoica-19,Hadjimichael-21}. 
All such phases present interesting functional properties such as particle-like behaviour~\cite{Aramberri-24}, ultrafast dynamics~\cite{Daranciang-12,Li-21,Gomez-24}, specific conductivity~\cite{Sharma-22}, chirality~\cite{Louis-12,Shafer-18} or negative capacitance~\cite{Iniguez-19} that justify their interest in nanoelectronic applications~\cite{Catalan-12,Salahuddin-2008,Parkin-08}. Moreover, the stabilization of one phase over the others is nowadays well understood and controlled by tuning the thicknesses of the ferroelectric and dielectric materials or the mechanical boundary conditions imposed by the substrate~\cite{Zubko-12,Hong-17,Dai-23}. 
Unfortunately, in most cases the resultant polarization pattern is fixed by the device shape and growing conditions limiting their tunability and hence restricting their practical use.

Due to the envisioned technologically relevant results, exploring alternative
strategies for an in-situ dynamical tailoring of the resulting polar topology is of critical importance.
The first steps towards this direction have already been taken by means of electric field or optical pulses~\cite{Daranciang-12,Li-19,Prosandeev-22,Prosandeev-23,Zajac-24} to pump polar optical phonons. In these works, different hidden phases could be stabilized by changing the frequency and shape of the pulses or by recurrently applying them a concrete number of times. While promising, these methods have so far stabilized only specific phases, and deterministic control of the entire phase diagram for topological textures remains elusive. Additionally, predicting the resulting polar arrangement given the shape and duration of the pulse is challenging.

Here, we propose a completely different and fresh perspective and focus on activating \emph{acoustic phonon excitations ({\sc{apex}})} to deterministically control the polar pattern in the material. This method exhibits more predictable behavior due to the strong coupling of ferroelectricity with both homogeneous~\cite{Choi-04,Dai-23} and inhomogeneous strains~\cite{Catalan-11,Lu-12,Zubko-13,Morozovska-21,Stengel-book}. In fact, the flexoelectricity and strong polarization-strain coupling are at the root of the effectiveness of mechanical conditions in stabilizing different complex phases. This is evidenced by the formation of complex topological structures in two-dimensional twisted freestanding layers~\cite{Bennett-23, Santolino-24}, around induced cracks~\cite{Shang-24}, on top of wrinkled surfaces~\cite{Kasai-24} and on rippled surfaces~\cite{Xu-24}. 
Unfortunately, in these studies, the mechanical boundary conditions are largely fixed and solely determined by the initial structural configuration i.e. the twist angle~\cite{Santolino-24}, the shape of the rippled surface~\cite{Xu-24}, the shape of the crack~\cite{Shang-24} or wrinkled surface~\cite{Kasai-24}.
Therefore, methods that rely on an on-demand externally tunable strain pattern remain challenging to achieve. 

In the present work, we fill this gap and demonstrate by means of atomistic second-principles~\cite{Wojdel2013,Escorihuela-Sayalero2017} simulations (details in Supporting Information) the deterministic control and stabilization of different polar textures at the bulk level in some model systems like BaTiO$_\mathrm{3}$ or PbTiO$_\mathrm{3}$ by activating {\sc{apex}} on the material. Besides, generalizations to other compounds showing ferroelectric modes are possible providing also a novel pathway to encounter new materials displaying non-trivial polar topologies. 

The Letter is organized as follows. First, we present the various domain structures that can be stabilized by pumping a unique {\sc{apex}}, depending on its direction and periodicity. Next, we explore the combination of several {\sc{apex}} to demonstrate the stabilization of more complex patterns, such as vortex-antivortex configurations or Skyrmion lattices. Finally, we show the recursive and dynamical control of the polar pattern, highlighting the deterministic back-and-forth transformation of the polar structure through the activation of different {\sc{apex}}.
\begin{figure}[bp]
    \centering
    \includegraphics[width=8cm]{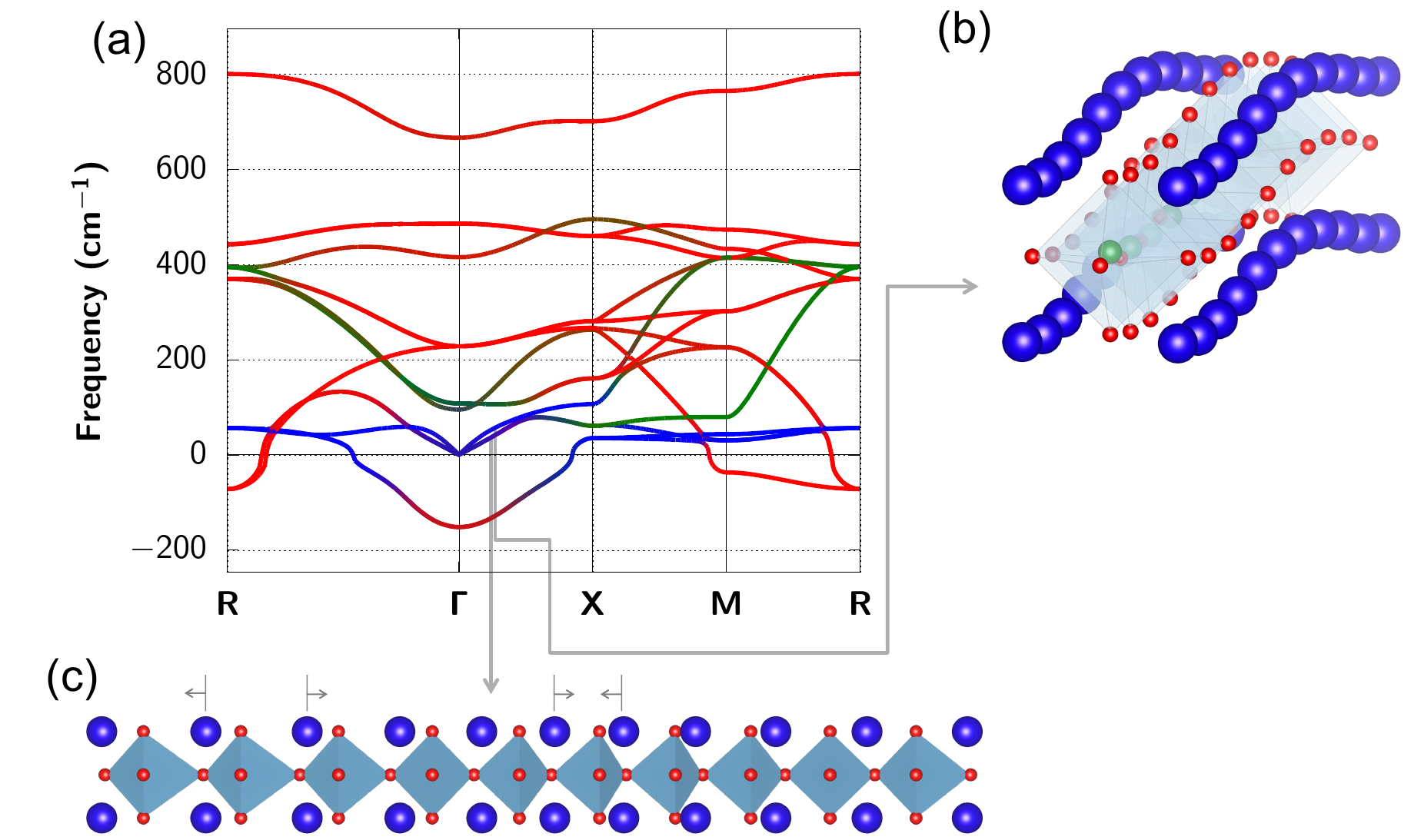}
    \caption{(a) Phonon dispersion curve of PbTiO$_\mathrm{3}$ in the chosen Pm$\bar{3}$m reference structure. The colors (Pb: blue; Ti: green; O: red) show the contribution of each atom to the eigenvectors. (b) Transversal {\sc{apex}}. (c) Longitudinal {\sc{apex}}.The amplitudes of the distortions have been augmented for visualization purposes.}
    \label{fig:phonon_disp}
\end{figure}

Before discussing the consequences on the polarization pattern, let us briefly examine the structural implications of activating {\sc{apex}} on the material. The results have been obtained starting from first-principles DFT results (PBEsol functional) and building on this second-principles models using {\sc{MULTIBINIT}} software~\cite{gonze2020abinit}. The second-principles models were then used to perform structural relaxations of the initial strain-gradients and molecular dynamics simulations. The acoustic phonon excitations were pumped using the {\sc{agate}} software~\cite{agate}. Further computational details can be found on the Supporting Information. In Fig.~\ref{fig:phonon_disp}, we show the phonon dispersion curve of PbTiO$_\mathrm{3}$ in the Pm$\bar{3}$m phase, as computed from first-principles at zero Kelvin. If we focus on the $\Gamma-X$ branch, we can distinguish between the doubly degenerate transversal mode [see Fig.~\ref{fig:phonon_disp}(b)] and the longitudinal mode [see Fig.~\ref{fig:phonon_disp}(c)]. The structural distortions generated by these {\sc{apex}} on the material will favour the development of head to head, or tail to tail domains in the latter, while in the former, they will lead to the formation of 180$^\circ$ domains. Indeed, similar cell deformations at domain walls are well explained from {\textit{ab initio}} calculations since the seminal work by Meyer and Vanderbilt~\cite{Meyer-02}. 
Having in mind the atomic deformation pattern resulting from the pumping of distinct {\sc{apex}}, we now demonstrate how strategically combining different modes and related cell deformations can promote the generation of various polarization patterns driven by flexoelectric coupling effects.

\begin{center}
\begin{figure}[btp!]
    \includegraphics[width=7.5cm]{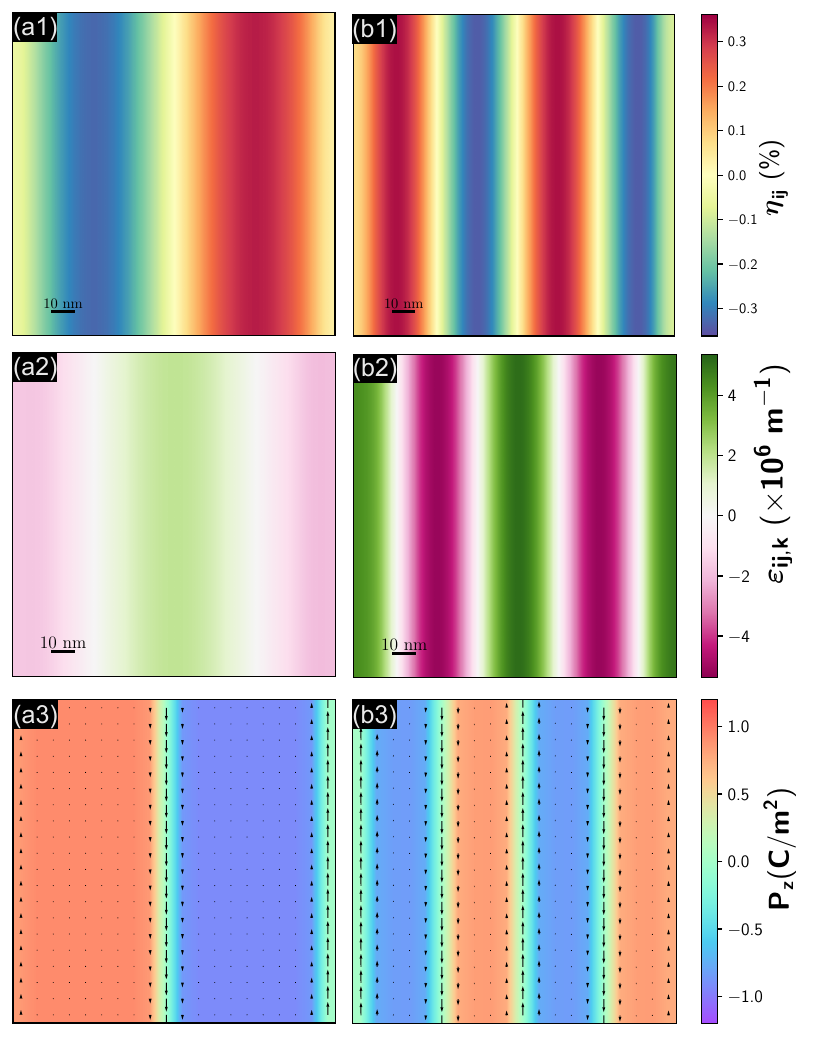}
    \caption{Stabilization of different domain walls in bulk PbTiO$_\mathrm{3}$ after the application of an {\sc{apex}} of amplitude $1$ \AA~and periodicities of $1/20$ and $1/10$ u.c.$^{-1}$. (a1-b1) Shear strain, $\eta_{xz}$, produced in the material. (a2-b2) Induced strain gradients, $\varepsilon_{xz,x}$. (a3-b3) Polarization patterns after relaxation of the supercell. Arrows indicate the in-plane components of the polarization whereas the color map represents the out of plane polarization.
    }
    \label{fig:Bloch_10_1_1}
\end{figure}
\end{center}

{\bf Single acoustic mode --}
We first focus on the case where a unique transverse {\sc{apex}} from the $\Gamma$ - $X$ acoustic branch is pumped into the Pm$\bar{3}$m phase of PbTiO$_\mathrm{3}$ inducing an overall distortion of $1$\AA~(which correspond to a maximal displacement of 0.015~\AA~of Pb atoms). Depending on its orientation, a significant strain gradient $\varepsilon_{xz,x}$ or $\varepsilon_{yz,y}$ is induced in the material as shown in Fig.~\ref{fig:Bloch_10_1_1}(a2-b2) and Fig. S1. Upon the relaxation of the supercell, this strain gradient induces the formation of ferroelectric domain walls~\cite{wojdel2014ferroelectric} through the flexoelectric effect as evidenced in Fig.~\ref{fig:Bloch_10_1_1}(a3-b3). Indeed, depending on the wavevector of the mode $q=1/20~{\rm u.c.}^{-1}$ or $q=1/10~{\rm u.c.}^{-1}$, where u.c. stands for unit cell, the magnitude of the strain gradient varies leading to initial induced polarization values of $1.14\ \mu$C/cm$^2$ and $3.88\ \mu$C/cm$^2$ respectively. After relaxation of the supercell, typical polarization  values of $95\ \mu$C/cm$^2$ and $84\ \mu$C/cm$^2$ respectively  (comparable to the bulk), are recovered inside the domain. 
The formation of a Bloch component within the domain wall arises from the ferroelectric nature of the domain walls~\cite{wojdel2014ferroelectric}, rather than being a direct result of the imposed strain gradient. The results presented are paradigmatic examples; however, other domain wall types, such as Ising or Néel, can also be engineered by pumping the longitudinal mode of the same phonon branch.
Besides, the amplitude and location of the strain and strain gradient are tunable by adjusting the amplitude and phase of the acoustic wave. As shown in Fig.~\ref{fig:Bloch_10_1_1}, the size of the domains can be designed by selecting different q-points from the acoustic branch. 
These domain walls are crucial for applications in non-volatile memory devices and advanced sensors~\cite{Catalan-12}, and their formation can be precisely controlled using acoustic phonon modes. Our study leverages the possibility to manipulate domain wall formation.
Remarkably, even if  they remain  stable  across the entire Brillouin zone, pumping a phonon from the acoustic branch with positive frequency, results in a phase that is lower in energy than the initial Pm$\mathrm{\bar{3}}$m configuration in which the phonon was pumped. 
%
This energy lowering can be explained by the coupling of the acoustic branch and the optical soft-mode~\cite{wang2019flexoelectricity,tagantsev1985theory,stengel2016unified,shirane1970soft} that after relaxation of the cell develops a non zero local polarization state.
{\bf Combination of acoustic modes --}
Up to now, it was shown that the location and size of domains can be controlled using a unique transverse {\sc{apex}}. However, it is also possible to pump simultaneously more than one phonon. Now, we demonstrate how mixing several {\sc{apex}} can create a multitude of combinations, leading to non-trivial polar textures in both BaTiO$_\mathrm{3}$ and PbTiO$_\mathrm{3}$.
\begin{center}
\begin{figure}[ht]
    \centering
    \includegraphics[width=9.2cm]{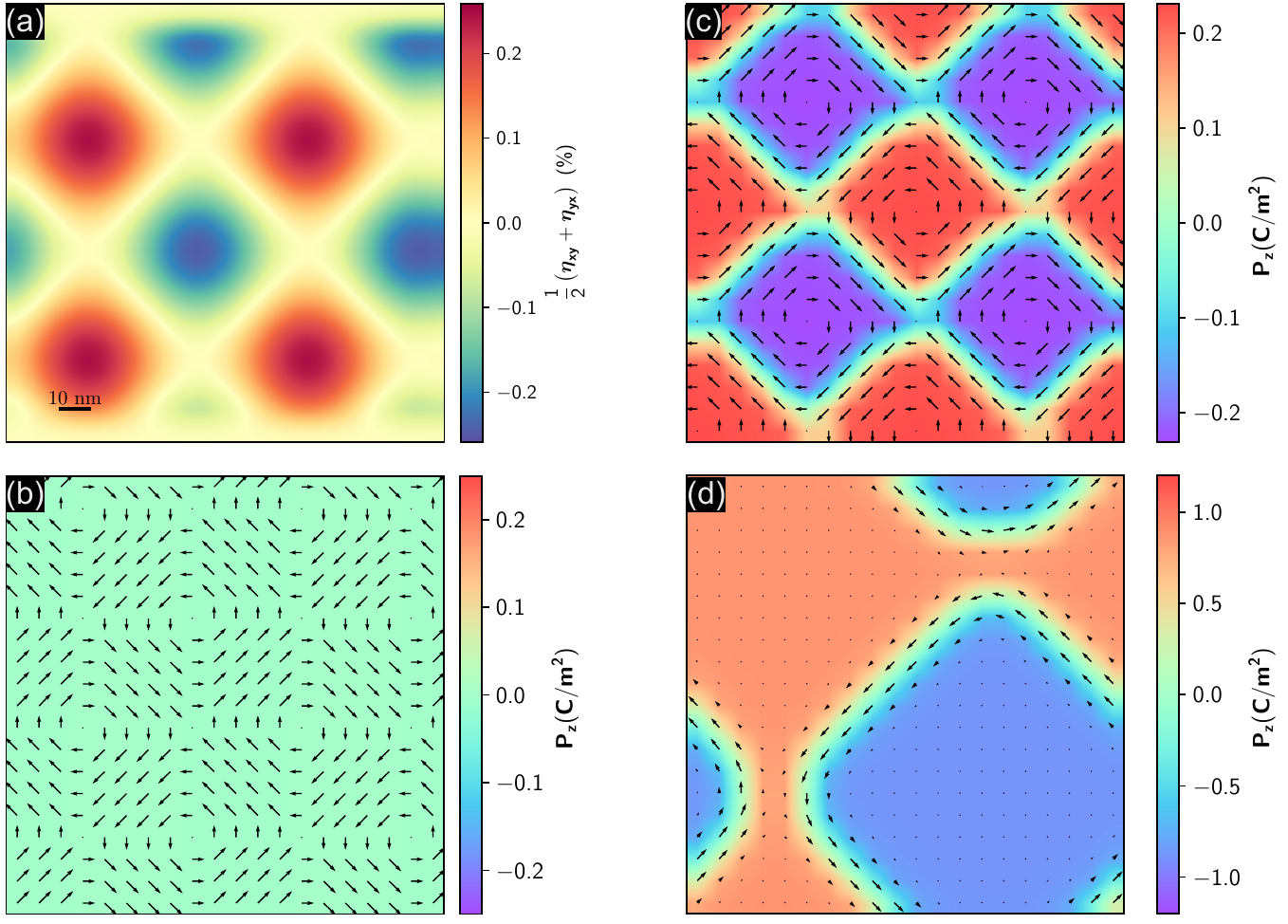}
    \caption{Stabilization of different topological defects by combining longitudinal and transversal acoustic phonons from the $\Gamma-X$ branch of the Pm$\bar{3}$m phase of (a-c) BaTiO$_3$ and (d) PbTiO$_3$.(a) Local initial in-plane shear strain, and (b) in-plane polarization after relaxation in BaTiO$_3$ by the pumping of transversal {\sc{apex}} along $x$ and $y$ with periodicities of $q=1/10$ u.c.$^{-1}$. (c) Polarization pattern after relaxation in BaTiO$_3$ displaying a meron-antimeron lattice when transversal {\sc{apex}} of periodicity $q=1/10$ u.c.$^{-1}$ are imposed on top of the in-plane distortion. For visualization purposes, the pattern has been shifted by 3 u.c. along the [110] direction relative to (a) and (b). (d) Polarization pattern after relaxation in PbTiO$_3$ displaying skyrmion defect as a result of transverse {\sc{apex}} with periodicities of $q^\prime=1/20$ u.c.$^{-1}$. Arrows indicate the direction of the in-plane components of the poalrization while color maps reflect the out-of-plane component.}
    \label{fig:2_modes}
\end{figure}
\end{center}

For instance, using a combination of longitudinal phonons from the $\Gamma$ - $X$ branch, an alternating pattern of positive and negative $\eta_{xy}$ shear strains, defined as $\frac{1}{2}\left(\frac{\partial u_y}{\partial x} + \frac{\partial u_x}{\partial y}\right)$, can be induced in bulk BaTiO$_\mathrm{3}$ where $u_x$ and $u_y$ are the atomic displacements in the $x$- and $y$-directions with respect to the undistorted cell respectively. Interestingly, this initial structure resembles the experimentally observed 2D ferroelectric vortex pattern in twisted BaTiO$_\mathrm{3}$ freestanding layers~\cite{Santolino-24}, as shown in Fig.~\ref{fig:2_modes}(a). Therefore, pumping this combination of longitudinal {\sc{apex}} subsequently gives rise, after relaxation, to an in-plane polar texture similar to that observed in twisted bilayers, as evidenced in Fig.~\ref{fig:2_modes}(b).
However, we can still go a bit further and combine the aforementioned modes with a transverse {\sc{apex}} similar to the ones discussed previously for the PbTiO$_\mathrm{3}$ case and create a non-zero $\eta_{xz}$ and $\eta_{yz}$ shear strain map. This induces the development of an out of plane component of the polarization on top of the in-plane components that leads to the formation of a meron/anti-meron lattice [see Fig.~\ref{fig:2_modes}(b)] showing semi-integer topological charge $Q$ as discussed in the Supporting information. This phase is similar to those already observed in PbTiO$_\mathrm{3}$/SrTiO$_\mathrm{3}$ freestanding layers~\cite{Shao2023} or chiral magnets~\cite{Yu-18}; however, is to the best of our knowledge, the first time that is predicted in BaTiO$_\mathrm{3}$.
Following a similar spirit of combining {\sc{apex}}, we can also try to stabilize more complex phases in PbTiO$_\mathrm{3}$. As shown previously, 
alternating positive and negative $\eta_{xz}$ and $\eta_{yz}$ strains can create ferroelectric domain walls. These ferroelectric domain walls have been demonstrated to be the key ingredient in nucleating polar skyrmions~\cite{Mauro-19}. By combining the transverse {\sc{apex}} that leads to ferroelectric domain walls along the $x$-axis with the mode that leads to ferroelectric domain walls along the $y$-axis, the creation of a polar skyrmion is naturally achieved, as shown in Fig.~\ref{fig:2_modes}(c). Details on the analysis of the topological charge can be found in the Supporting information.

As demonstrated by the previous examples, fine-tuning of the polarization texture can be achieved in various ferroelectric materials by combining several {\sc{apex}}. The discussed method not only proves effective in stabilizing previously reported polarization textures but also predicts novel configurations. Furthermore, the method is deterministic, allowing the resulting polarization pattern to be anticipated a priori after selecting the appropriate modes.

{\bf Dynamical control of polar patterns --}
Until now, we have focused solely on the stabilization of various polar patterns. However, unlike conventional methods where mechanical boundary conditions are solely fixed by growth conditions, the flexoelectric response to acoustic waves is {\it dynamical}~\cite{wang2019flexoelectricity}, enabling real-time manipulation of polar textures by activating {\sc{apex}}.

This section provides a concrete example in demonstrating that our method enables dynamical control of the orientation of the ferroelectric domains in PbTiO$_\mathrm{3}$. 
Previous studies have shown that the orientation of the domains exhibits a stochastic Vogel-Fulcher-type behavior~\cite{Gomez-24}. 
\begin{center}
\begin{figure*}[!]
    \centering
    \includegraphics[width=\textwidth]{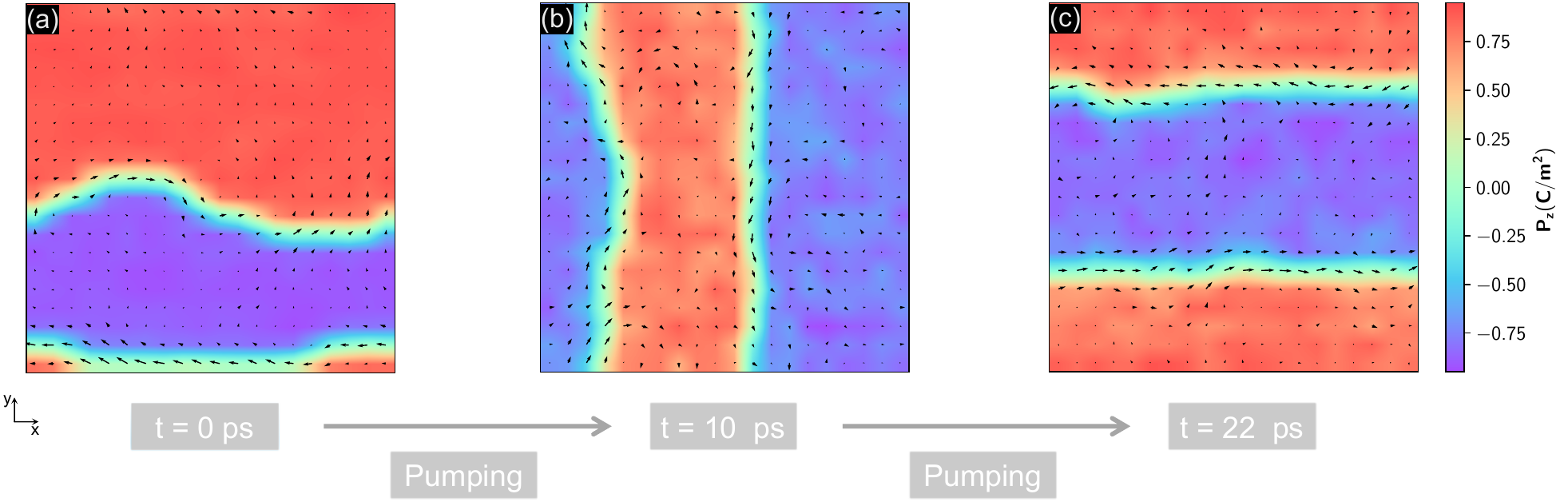}
    \caption{Dynamical tuning of domain orientation in PbTiO$_\mathrm{3}$ by {\sc{apex}} as described in the main text. Snapshots of molecular dynamics simulation at (a) $0$ ps showing a domain along the $x$-direction, (b) $10$ ps showing a domain along the $y$-direction after the application of the acoustic phonons and (c) $22$ ps showing the reversal to a $x$-oriented domain driven by the phonon pumping. Arrows indicate the in-plane components of the polarization whereas the color map represents the out of plane polarization.}
    \label{fig:dynamics}
\end{figure*}
\end{center}
Relying on the specific conductive properties of domain walls~\cite{Sharma-22}, a deterministic control of this orientation might so reveal crucial for designing novel nanoelectronic devices such as transistors~\cite{ou2023ferroelectric}.

To achieve precise control over the orientation of the ferroelectric domains, we employ a two-step process. Initially, a transverse {\sc{apex}} with a periodicity of $q_1=1/10~{\rm u.c.}^{-1}$ is pumped. The strain gradient associated to this periodicity, of approximately $6.1\times 10^7~{\rm m}^{-1}$ has been shown to be sufficient to reorient the domains with a moderate phonon amplitude of 12~\AA. This corresponds to a maximal lead displacement of 0.17~\AA\ compared to the cubic phase that do not break the stability of the crystal. Subsequently, we pump {\sc{apex}} with the desired periodicity, specifically $q_2=1/20~{\rm u.c.}^{-1}$ in this instance. This dynamical approach underscores the potential of our method for fine-tuning polar textures in advanced material applications.
As illustrated in Fig.~\ref{fig:dynamics}(a), we begin the simulation with the ferroelectric domains oriented along the $x$-direction. We then pump $q_1$ eight times at intervals of $0.36$ ps, propagating along the $x$-direction. Note that, as shown in Fig.~\ref{fig:Bloch_10_1_1}, these acoustic phonons along the $x$-direction would favour the $y$-orientation of the domains.
Afterwards, $q_2$ is pumped four times at intervals of $0.36$~ps, again propagating along the $x$-direction. Finally, the system is allowed to evolve over a period of $7$~ps and as shown in Fig.~\ref{fig:dynamics}(b) the domain orientation is flipped. To revert again the domain orientation and transit back to an $x$-oriented domains [see Fig.~\ref{fig:dynamics}(c)], the same procedure is applied but now with the wave propagating along the $y$-direction.

As demonstrated here, the orientation switching of the domains can occur even at relatively low temperatures where the domains are typically expected to be static~\cite{Gomez-24}. This suggests that effective tuning of the polar phase by means of {\sc{apex}} could be achieved at room temperature, making it a promising candidate for new electronic applications. Moreover, this example serves as a proof of concept, indicating that this dynamical control approach could be extended to other phenomena, thereby offering new opportunities for the control of polar patterns. Future work will explore these potential applications in greater detail.

In this work we have presented a novel method for the deterministic control of polar textures in various ferroelectric materials through {\sc{apex}}. By carefully selecting and combining longitudinal and transverse acoustic phonons, various topological defects in BaTiO$_\mathrm{3}$ and PbTiO$_\mathrm{3}$ have been induced, demonstrating a high degree of control over the resultant polarization patterns. This novel method offers significant advantages over traditional pumping of polar optical modes from electric field or optical pulse techniques, as it provides a more predictable behavior and a greater tunability of the phase diagram. The ability to dynamically manipulate polar textures in situ holds promise for the development of next-generation nanoelectronic devices, where precise control over ferroelectric phases is crucial.
Moreover, while this work has concentrated on the fine-tuning of polar topologies, the method is versatile and can be extended to stabilize hidden phases in other systems of interest showing a complex energy landscape, like HfO$_2$ or antiferroelectric PbZrO$_3$. Moreover, we have only investigated excitations of acoustic standing waves. Exploring further the behavior of propagating waves within the material could also provide valuable new opportunities for the displacement of polar nanodomains.

From an experimental standpoint, the quasi-monochromatic generation and control of acoustic phonons in superlattices of perovskite oxides have already been demonstrated~\cite{Ng-22,Yang-21} with wavelengths of  the order of tens of nanometers which perfectly match our proposed values. Furthermore, resonant ultrasound spectroscopy is routinely employed to measure the elastic and dielectric constants of piezoelectric materials~\cite{Balkirev-19}. These developments suggest that activating {\sc{apex}} in materials is feasible, thereby supporting the experimental viability of our proposal. This finding also strongly motivates further development of practical setups for generating controlled acoustic pulses.
Our findings pave the way for further theoretical and experimental research on using {\sc{apex}} as a new promising avenue to design and optimize functional ferroelectric materials.
\begin{acknowledgements}
Authors acknowledge P. Zubko, A. Caviglia, P.-E. Janolin and J.-M. Triscone for their insightful discussions.
L.B, F.G.-O. and Ph. G. acknowledge support by the European Union’s Horizon 2020 research and innovation program under Grant Agreement No. 964931 (TSAR).
The authors acknowledge the use of the CECI supercomputer facilities funded by the F.R.S-FNRS (Grant No. 2.5020.1) and of the Tier-1 supercomputer of the Fédération Wallonie-Bruxelles funded by the Walloon Region (Grant No. 1117545). 
F.G.O. also acknowledges financial support from MSCA-PF 101148906 funded by the European Union and the Fonds de la Recherche Scientifique (FNRS) through the grant FNRS-CR 1.B.227.25F. Ph. G. also acknowledges support from the Fonds de la Recherche Scientifique (FNRS) through the PDR project PROMOSPAN (Grant No. T.0107.20).
S. A. acknowledges financial support from the EIT Raw Materials - AMIS Joint Master Program
\end{acknowledgements}

\end{document}